# Observation of *c*-axis Magnetization at Low Temperatures in Weak Ferromagnet FeBO$_3$ Reveals a Spin-Reorientation Transition

J. Franklin,[1] J. Pfund,[1] J. Bedard,[1] W. Zhang,[2] P. Shiv Halasyamani,[2] M. Jain,[1,3] I. Sochnikov,[1,3,4]

[1]*Department of Physics, University of Connecticut*
[2]*Department of Chemistry, University of Houston*
[3]*Institute of Material Science, University of Connecticut*
[4]*Material Sciences & Engineering Department, University of Connecticut*

The weak ferromagnet $FeBO_3$ is well known for being a unique system for modelling and testing magnetic dynamics primarily due to relatively simple and localized magnetic structure and its interesting spin wave dynamics. At room temperature, it has slightly canted iron moments lying in the *a-b* plane that result in a strong antiferromagnetic moment and a weak ferromagnetic moment, which results in pronounced ferromagnetic and antiferromagnetic spin modes. However, some previous studies have shown unusual low-temperature behavior that suggests a phase transition. By performing low-temperature magnetization measurements, both in bulk and on the mesoscale, we have observed a low temperature magnetic texture in this material in which a large *c*-axis magnetization occurs. Magnetic fields along the *c*-axis as high as 1300 Oe were observed close to the sample's surface. This presents evidence for the onset of a Morin transition or another type of spin-reorientation phase transition wherein the $Fe^{3+}$ moments would acquire a *c*-axis component to their canting below a critical temperature. The observation of this *c*-axis magnetization suggests that there is a different ground state in this material than has been previously expected and could be due to as yet unexplored intricacies of the Dzyaloshinskii-Moriya interaction.

**Introduction**

Iron borate has been the subject of research attention for decades due to its unique magnetic properties. Historically, weak ferromagnets such as iron borate played a key role in developing the theory of Dzyaloshinskii-Moriya (DM) interactions [1], and even today iron borate and related materials are an exciting environment to study DM interactions [2,3] and their consequences, such as the magnetoelectric effect [4] and skyrmions [5]. In addition, iron borate has already been proven to be a valuable material for studying effects in magneto-optics [6,7] and magnetism at high pressures [8–11].

Iron borate, $FeBO_3$ (FBO), is an antiferromagnet with weak ferromagnetism and has a Néel temperature of 348 K. It has a calcite-type crystal structure with point group symmetry $D_{3d}$ and space group $D^6_{3d}$. It can be described by either rhombohedral or hexagonal lattice vectors [12] but in this work, reference to lattice vectors will always be to the hexagonal unit cell, e.g., the *c*-axis is in the hexagonal (0 0 1) crystallographic direction, see Figure 1. In each unit cell, there are two iron sites. At room temperature the moments of these iron sites lie in the easy plane of magnetization, which is the *a-b* plane. These moments are mostly antiparallel, giving rise to the strong antiferromagnetism, but have a small (~1°) in-plane canting, which gives rise to a weak parallel component and thus a weak in-plane ferromagnetism [13], see inset of Figure 1 (a). This coexistence of canted antiferromagnetism and ferromagnetism (canted AFM/FM) is what makes possible the pronounced spin wave modes in this material [11,14,15].

While research on iron borate is rich, characterization at low temperatures is sparse. In a few of the works on $FeBO_3$ that went as low as 4 K, surprising results were found. For example, Svistov *et al.* [16] measured the change in magnetization of iron borate due to microwave pumping and found that at low temperatures, the change in magnetization was opposite that in high temperatures. In other works, Seleznyova [17,18] performed EMR measurements on iron borate and gallium doped iron borate and found that the gallium doped samples, which have the same crystal and magnetic structure as pure iron borate, had antiferromagnetic resonance (AFMR) peaks which had an unexpected change in their temperature dependence at low T. Specifically, these AFMR peaks shifted to lower fields from the Néel temperature until some low T (<100 K), and then started shifting back to higher fields until the lowest measured T of 4 K. Seleznyova suggested that this behavior was reminiscent of hematite, which undergoes a Morin (or spin-flop) transition at low temperatures [19–21], and suggested that low temperature magnetization measurements be performed.

In the present work, we report magnetization measurements of $FeBO_3$ from above room temperature to as low as 50 mK, both in bulk and on the mesoscale. We observed *c*-axis magnetization that could be explained by the onset of a Morin-type transition that causes the iron moments to rotate and acquire a component along the *c*-axis, see Figure 1 (b).

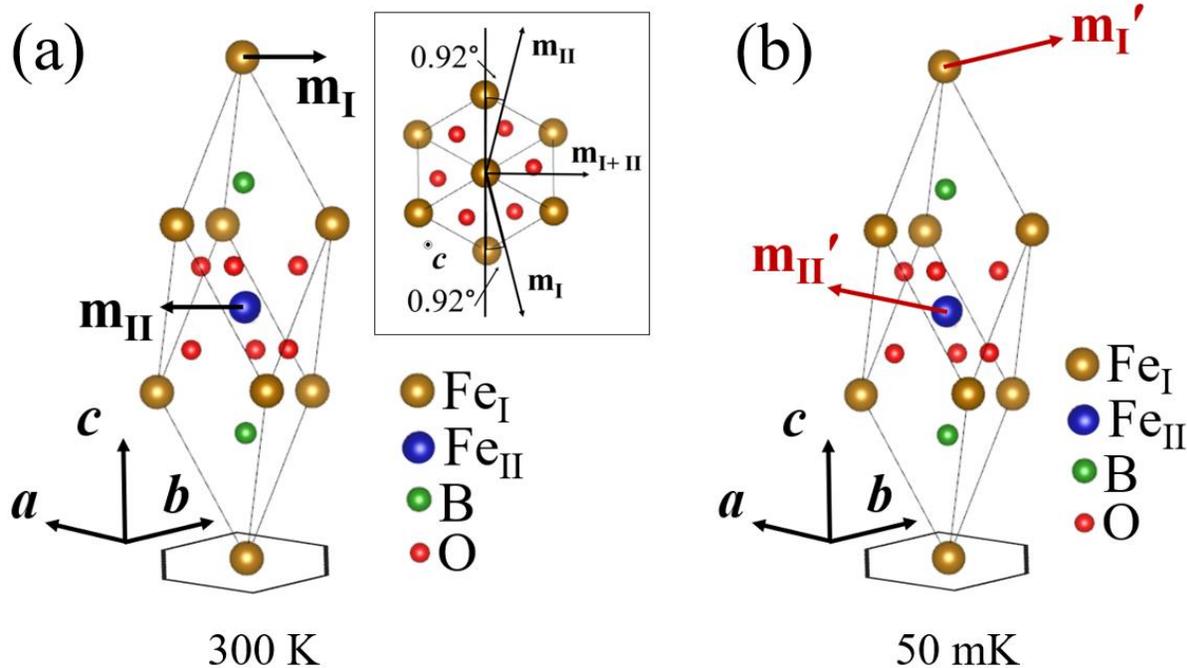

Figure 1. Room temperature versus low temperature magnetic structure of $FeBO_3$. The crystal structure of $FeBO_3$ is rhombohedral calcite-type as seen in panel (a). The axes shown are the hexagonal unit vectors, and relative to these the easy plane of magnetization is the *a-b* plane. The iron atoms are located at two sites, denoted by $Fe_I$ and $Fe_{II}$. Their moments, shown by $m_I$ and $m_{II}$, lie in the *a-b* plane. The inset of (a) shows the unit cell looking down the *c*-axis, revealing the in-plane room temperature magnetic structure. The moments of the two iron sites are mainly antiparallel but have a slight canting in-plane (0.92°), resulting in a strong antiferromagnetic moment and a weak ferromagnetic moment, denoted by $m_{I+II}$. Our results suggest that at low temperature, the moments become canted in the c direction, shown in red by $m_I'$ and $m_{II}'$ in (b).

**Experiment**

$FeBO_3$ crystals were grown with reagents $Fe_2O_3$ (Alfa Aesar, 98%, 6.3871 g), $H_3BO_3$ (Alfa Aesar, 99.8%, 30.9165 g), and $Bi_2O_3$ (Alfa Aesar, 99%, 27.9602 g) by the top-seeded solution growth (TSSG) method. $Fe_2O_3$, $H_3BO_3$ and $Bi_2O_3$ compounds with a molar ratio 2:25:3 were mixed, placed in an Ø40×40 $mm^3$ platinum crucible, and slowly heated to 1050 °C to melt the reagents. The melt was held at 1050 °C for 20 hours to ensure homogeneity. The temperature was then quickly cooled, over 20 minutes, to 789 °C, the crystal growth temperature. A seed $FeBO_3$ crystal was introduced to the melt and subsequently rotated at 10 rpm. The temperature was decreased slowly, by 0.2 °C /day, until crystals had grown on the seed. When the crystal growth was completed (determined by observing the crystal size in the melt), the as-grown crystals were pulled out from the melt and held at 5 mm above the melt surface. The furnace was cooled down to room temperature at a cooling rate of 20 °C/hour. At room temperature, the as-grown crystal was taken out of the furnace. The entire crystal growth process took about 30 days. The crystal structure and phase purity of this FBO single crystal was examined by X-ray powder diffractometer (XRD; Bruker D2 Phaser) with Cu Kα radiation (λ = 1.54 Å). See Appendix A for the XRD data with peaks close to 37.5 and 79.6 degrees, corresponding to (0 0 6) and (0 0 12) planes in the hexagonal system affirming the single-phase and (0 0 1) orientation of the present FBO single crystals.

Bulk magnetization measurements were performed using an Evercool Physical Property Measurement System (PPMS) from Quantum Design Inc. with a vibrating sample magnetometer (VSM) attachment. To measure the out-of-plane magnetization, the crystal was mounted so that the *c*-axis of the crystal was parallel with the applied dc magnetic field of 4 T. The sample was first heated to 400 K before performing zero-field cooled (ZFC) and field-cooled (FC) measurements in succession with 4 T field. Data was collected over a temperature range of 3 - 400 K, with steps of 2 K and 0.5 K for the ranges T < 330 K and T > 330 K, respectively.

Scanning Superconducting QUantum Interference Device (SQUID) measurements were performed in a Bluefors LD250 dilution refrigerator. A SQUID is a highly sensitive magnetic field to voltage transducer [22–25]. This quantum sensing technique allows for immense sensitivity, and thus the field coils can be made on the order of micrometers. For the measurements in this work, the pickup loop of the SQUID had a radius of 400 nm. Using this microscope, images are generated by scanning the SQUID in a raster fashion parallel to the surface of the sample using our lab-built piezoelectric microscope [26–29]. A typical distance of the pickup loop above the sample is 3 μm for the measurements in this work. Since the $FeBO_3$ samples studied here are grown with the (0 0 1) crystallographic direction, or *c*-axis, normal to the bulk sample surface, the raster SQUID images in this work show (0 0 1) magnetization (out-of-plane) with spatial resolution on the mesoscale. See Appendix B for more details on the scanning SQUID apparatus.

**Results**

Temperature dependent dc magnetization M(*T*) data of the $FeBO_3$ single crystal in ZFC and FC modes with 4 T field applied parallel to the crystal's *c*-axis is shown in Figure 2. The abrupt transition at 347 K is identified as the Néel temperature, $T_N$ marked as (1). As can be seen in Figure 2, the magnetic moment does not reduce to zero even at temperatures higher than $T_N$. This is because even in the paramagnetic region, there is a high moment due to the applied field. This also explains the overall slope in the ZFC data between $T_N$ and 400 K. A slight downward shift in the FC data close to 250 K could be due to a slight shift in the sample during measurements as the crystal was wrapped in Teflon before loading it into the VSM brass sample holder.

The magnetization values jump dramatically at both 46 K and 25 K, separated by a slightly positive slope, indicating two distinct transitions. The 25 K transition is also noted to have a significantly larger slope than its 46 K counterpart, which suggests that the two transitions are not of the same nature, but rather originate from two different sources. ZFC and FC signals split below 46 K, which is indicative of the emergence of *c*-axis FM ordering and should correspond to hysteresis in the *c*-axis magnetization in that temperature range. This low temperature upturn also increases in slope with decreasing temperature, indicating that this transition is still incomplete at 3 K.

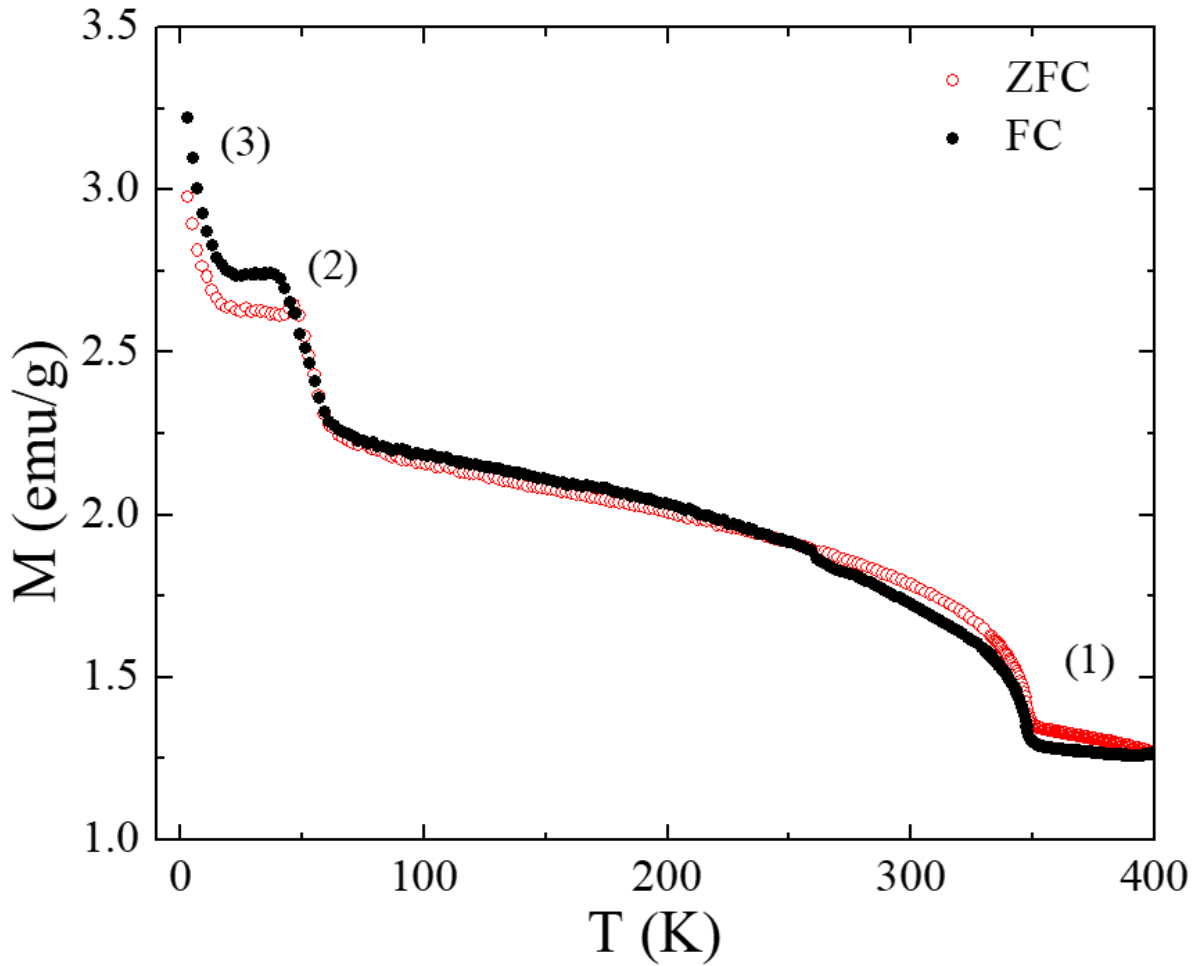

Figure 2. Out-of-plane temperature dependent magnetization data of bulk single crystal $FeBO_3$ in Zero Field Cooled (ZFC) and Field Cooled (FC) modes with applied field of 4 T. Three main features are noted, labeled (1), (2), and (3). The sample's Neel temperature, 347 K, is seen at (1). An unknown feature is observed at (2), in the temperature range of 25 K – 46 K. (3) shows a clear increase in magnetization with decreasing temperature for the temperature region 3 K – 25 K.

Figure 3 shows magnetic images from scanning SQUID measurements. Figure 3 (a), (b), and (c) show images at three different locations on the sample. In these images a dark red indicates a strong magnetization oriented out of the sample and a dark blue indicates a strong magnetization into the sample (both along the *c*-axis). Since these samples are grown with the *a-b* crystallographic plane as the sample surface, these images show the strength of magnetic field components parallel to the *c*-axis. The images show in all locations magnetized domains with sizes on the order of tens of microns. The magnitude of the fields measured above these domains is on the order of 500-1300 Oe, as seen in the cross sections plotted in panel (d) of Figure 3. In panel (c), the image was taken at the edge of the sample, and the tan line indicates the approximate sample edge. The region with a positive field is a magnetized domain on the sample, whereas the region with negative magnetic field is off the sample. This off-sample field is a distant field from the positive moments on the sample. The magnitude of the field in this image is notably smaller than the field over the domains in the middle of the sample. The only post-processing performed on this data is the offset of a few sections of lines in each image where the SQUID performed a flux jump, which is the sudden change in the signal by an integer number of flux quanta [30]. Thus, the strong *c*-axis magnetic fields in these measurements are essentially raw data.

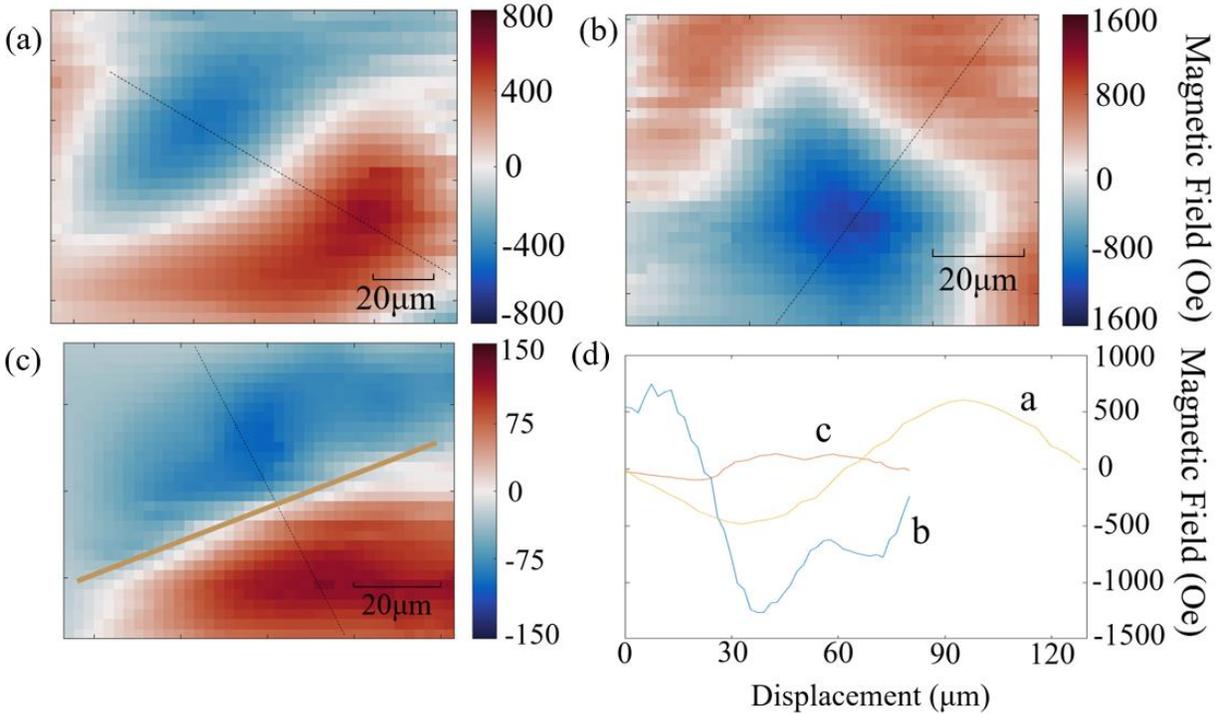

Figure 3. Magnetic fields above an *a-b* surface of FeBO$_3$. (a), (b), and (c), show magnetic images at different locations on the sample, all at ~50 mK. The image in (c) was taken at the edge of the sample, and the solid tan line represents the approximate sample edge. The red region in (c) is a magnetized domain on the sample, and the blue region is an area of stray field off the sample. (d) shows cross sections a, b, and c that are represented by the dotted lines in (a), (b), and (c), respectively. These images show *c*-axis magnetic fields that are on the order of 500-1300 Oe. Since the room temperature magnetic structure of FeBO$_3$ contains only moments lying in the *a-b* plane, any component aligned with the *c*-axis is unexpected.

## Discussion

Using a model published by Craik [31], we calculated the effect that out-of-plane canting of the iron moments in the FeBO$_3$ unit cell would have on magnetization near the surface. We found that 600 Oe, a typical magnetic field magnitude from our results, would correspond to ~5° of out-of-plane canting. This is a conservative estimate, however, because fields as strong as ~1300 Oe were observed. Details of this estimate are provided in Appendix C.

Previous studies on $FeBO_3$ to our knowledge were performed only at higher temperatures (T ≥ 4 K) and found only weak, in-plane magnetization of about 115 Oe [32–34]. In order to explain our observation of *c*-axis magnetization, we suggest that the $Fe^{3+}$ moments may be experiencing the onset of a Morin-like transition, wherein the moments change direction under some critical temperature, see Figure 1 (b). In hematite, the magnetic state at temperatures below the Morin transition is a perfect antiferromagnet with the $Fe^{3+}$ moments aligned with the hexagonal *c*-axis. This state has no bulk *c*-axis magnetization, but the transition between the two states is complex and intermediate states that have *c*-axis bulk magnetization may occur [35]. Iron borate is isomorphic with hematite, and our results could be explained by an incomplete Morin transition, wherein some intermediate states towards that transition are observed. It is possible that the final state of the proposed transition is the ground state of this system at some nonzero temperature, fully aligning the $Fe^{3+}$ moments antiferromagnetically with the *c*-axis, or that it is suppressed to 0 K. Alternatively, the transition could be a different type of spin-reorientation transition [36] which reaches a ground state where the $Fe^{3+}$ moments align with the *c*-axis ferromagnetically or reach some maximum canting angle. Yet another possibility is that this material underwent a strain-induced transition. It has been observed that the Morin transition temperature is enhanced with hydrostatic pressure in hematite [37] and that a Morin transition can be induced in iron borate under pressure [38,39]. No strain was intentionally applied to our samples, but it is possible that iron borate at low temperatures has a ground state which is highly sensitive to small strains. Our scanning SQUID sample was mounted with a small drop of varnish far away from the measured region in order to mitigate strains from thermal expansion. For bulk magnetization measurements the sample was not glued but wrapped in Teflon tape, thus the role of strains should be minimal in those measurements. Overall, we find it possible but unlikely that small external strains could have arisen and caused the *c*-axis magnetization. Further theoretical work expanding upon previous theoretical calculations [40] and focusing on these potential effects would be desired. Whatever the case, the unexpected canting indicates that there may be an unexplored evolution of the DM interaction in the ground state of this material, and this in turn may have rich consequences on the dynamics of the system, such as the spin wave modes.

**Conclusions**

Previous studies that hinted towards interesting magnetic behaviors in $FeBO_3$ at low temperatures have been strongly supported with the observation of a new magnetic phase with both bulk and scanning SQUID magnetization measurements. This phase is characterized by strong *c*-axis magnetization at low temperatures, and we interpret these results as the onset of a Morin-like transition. The presence of this transition implies that the ground state in this system is not the expected *a-b* plane canted AFM/FM, the same as the room temperature phase. Instead, our results show that a first-principles description of the ground state of this material remains to be fully developed. Perhaps this will have implications for other canted AFM/FM materials wherein the DM interaction dominates the magnetic structure.

**Acknowledgements**

JF, JP, JB, MJ, and IS gratefully acknowledge the Air Force Research Laboratory, Materials and Manufacturing Directorate (AFRL/RXMS) for support via Contract No. FA8650 – 21– C5711. JF, JP, JB, MJ, and IS also thank S. Pamir Alpay, Sanjeev Nayak, and Serzat Safaltin for fruitful discussions and feedback. WZ and PSH thank the Welch Foundation (Grant E-1457) and Azimuth / AFRL (Contract No. 238-5404-UH2) for support.

**Appendix A: X-Ray Diffraction Data**

The crystal structure and phase purity of the FBO single crystals was examined by X-ray powder diffractometer (XRD; Bruker D2 Phaser) with Cu Kα radiation (λ = 1.54 Å). Figure A1 shows the XRD scan of the sample used for scanning SQUID with peaks close to 37.5 and 79.6 degrees, corresponding to (0 0 6) and (0 0 12) planes in the hexagonal system affirming the single-phase and (0 0 1) orientation of the present FBO single crystals. The scan on the sample used for the bulk measurements resulted in a nearly identical spectrum, which is expected because both samples were acquired from the same source.

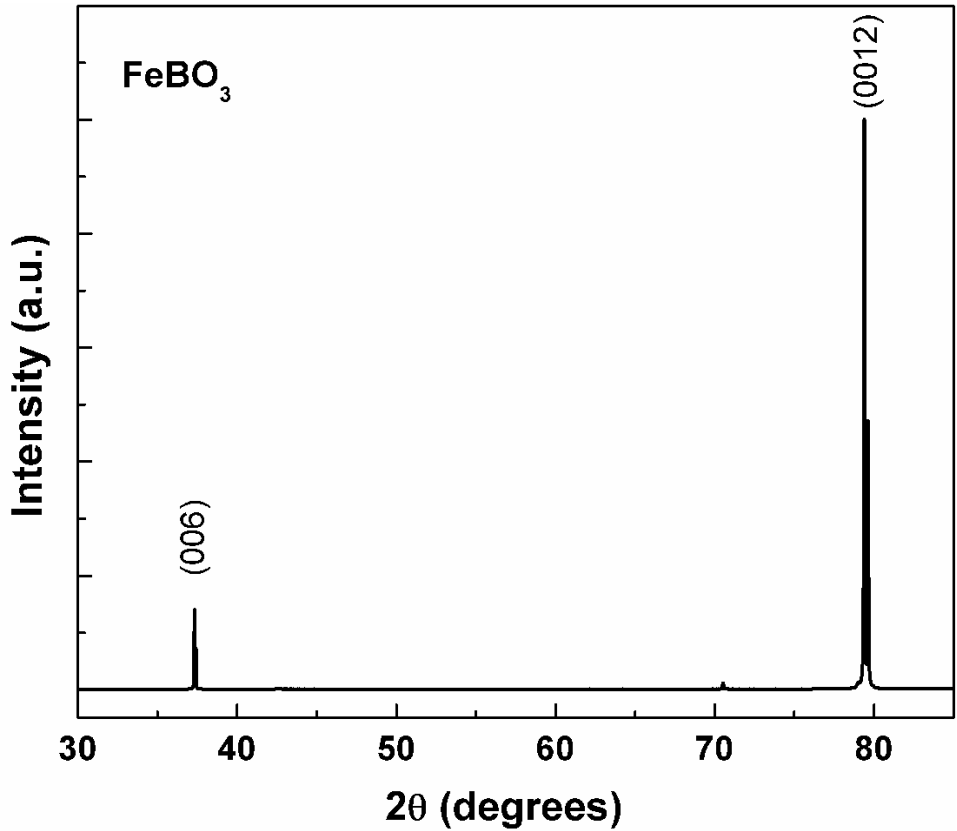

Figure A1. X-Ray diffraction results. Shown is the XRD spectrum resulting from a scan in a powder diffractometer. The presence of only the (0 0 6) and (0 0 12) peaks confirm that only the hexagonal *c*-axis has planes which are parallel to the surface of the sample.

**Appendix B: Scanning SQUID Apparatus**

A SQUID is a highly sensitive magnetic-flux-to-voltage transducer. A modern SQUID can be thought of as a miniaturized version of typical bulk mutual inductance loops, see Figure A2 (a). Uniquely, however, a SQUID utilizes the electron wavefunction interference between two superconducting regions in a structure known as a Josephson Junction, see Figure A2 (b). These Josephson junctions are where the immense field sensitivity of a SQUID is derived from.

Our SQUID is positioned by a piezoelectric scanner and is scanned across the surface of a sample in a raster fashion to measure magnetic images, see Figure A2 (c). Since the SQUID is aligned with the pickup loop parallel to the sample's surface, the flux that passes through the pickup loop will consist only of contributions from out-of-plane components of magnetic fields generated by the nearby sample. Thus, a SQUID that is aligned in this way will measure the out-of-plane component of magnetic fields from the sample.

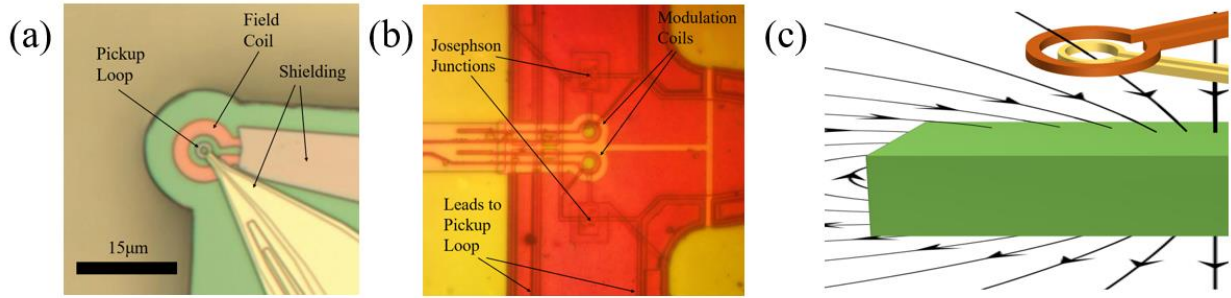

Figure A2. Scanning SQUID. In panel (a), a photograph of a SQUID is shown with the field coil, pickup loop, and shields labelled. The pickup loop and field coil are positioned on the tip of the SQUID chip so that they can be positioned as close to the sample as possible. The field coil is added to most SQUIDs to provide small auxiliary fields to the region around the pickup loop and add the capability to measure susceptibility. Magnetic flux that passes through the pickup loop is converted to voltage by the Josephson Junctions in the center of the chip, which are shown in panel (b) along with the modulation coils. Panel (c) shows a schematic of a SQUID measuring the magnetic field above a magnetized sample. The SQUID is represented by the orange field coil and yellow pickup loop, and the sample is the green solid. Magnetic field lines of an out-of-plane magnetic dipole are shown to represent a sample that is magnetized out-of-plane. When the SQUID is scanned over the sample, the pickup loop measures the magnetic fields caused by the nearby magnetized sample. We scan the SQUID over the sample surface to measure magnetic images and resolve magnetic landscapes on the mesoscale.

**Appendix C: Canting Angle Calculation**

Craik [31] calculates the components of the magnetic field above a domain. Craik's result for the out-of-plane component of the magnetic field is given by Equation A1.

$$H_z = \sigma \sin^{-1}\left(\frac{(c+a)b}{((c^2+a^2)+z^2)^{1/2}(b^2+z^2)^{1/2}}\right) - \sigma \sin^{-1}\left(\frac{bc}{(c^2+z^2)^{1/2}(b^2+z^2)^{1/2}}\right) \quad \text{Equation A1}$$

In Equation A1, $H_z$ is the out of plane component of the magnetic field, $\sigma$ is the magnetization of the domain (in c.g.s.), $c$ is the distance from the point below the point of measurement to the domain, $a$ and $b$ are the two dimensions of the rectangular domain, and $z$ is the distance the point of measurement is above the surface of the sample, see Figure A3.

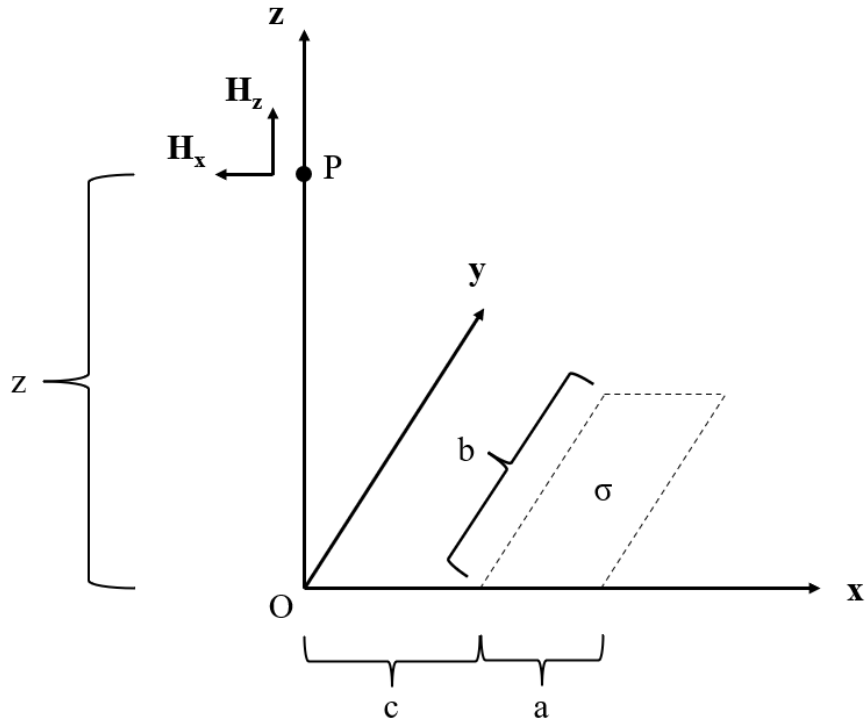

Figure A3. Set up for the calculation of the magnetic field above a magnetized domain. The field components $H_x$ and $H_z$ are calculated by Craik at a point P which is a distance z above the origin O. A nearby rectangular domain lies in the **x-y** plane with one edge on the **x** axis and is magnetized with magnetization σ (positive magnetization is parallel to the positive **z** direction). The domain is a distance c from the origin to the nearest edge along the **x** axis and it has an x dimension a and a y dimension b. In this calculation, c may be set negative to position the domain directly below point P, and rectangular domains of arbitrary position (not with an edge along the **x** axis) can be considered by summing the field contributions from multiple domains with **x** axis edges.

We assume a square domain with sides of 40 μm and a height over the sample of 3 μm. To find the field contribution above the center, we will take the field contribution of a half-square and then double it. Thus, $a = 40$ μm, $b = 20$ μm, $c = -20$ μm, $z = 3$ μm. Using these assumptions, the result of Equation A1 is, after doubling, $H_z = \sigma(5.44)$.

To solve for the magnetization of iron borate, we will start with the moment of the Iron atoms and use elementary quantum mechanics [41]. We assume that the spin moments are dominant, and the spin moment of an atom is given by Equation A2, where n is the number of unpaired valence electrons.

$$\mu = \mu_B \sqrt{n(n+2)} \qquad \text{Equation A2}$$

Iron has 5 unpaired valence electrons, resulting in $\mu = 5.91\ \mu_B$. We assume the crystal structure of iron borate remains the same, and thus iron borate has a unit cell volume of $V = 89.51$ Å$^3$ [42]. The magnetization (moment per unit volume) is then

$$\sigma = \frac{5.91\mu_B}{89.51\text{Å}^3}\sin\theta \cdot \frac{9.27*10^{-24}\,Am^2}{1\mu_B} \cdot \frac{1*10^{30}\,\text{Å}^3}{1m^3} \cdot \frac{1Oe}{1000Am^{-1}} = 612\sin\theta\,Oe,$$

where $\Theta$ is the out-of-plane canting angle of the iron moments. We double the result because there are two irons per unit cell in FBO, resulting in $\sigma = 1224\sin\theta\,Oe$. So, using our result that $H_z = \sigma(5.44)$, we have $H_z = 6659\sin\theta$ Oe. Using this result, our findings of ~600 G would correspond to ~5 degrees of out-of-plane canting.